\documentclass[runningheads]{cl2emult}

\usepackage{makeidx}  
\usepackage{graphicx} 
\usepackage{subeqnar} 
\usepackage{multicol} 
\usepackage{cropmark} 
\usepackage{lnp}      
\makeindex            

\newcommand{\euler}[1]{{\usefont{U}{eur}{m}{n}#1}}

\newcommand{\umu}{\mbox{\euler{\char22}}}

\newcommand{\mnras}{\mbox{MNRAS}}
\newcommand{\apj}{\mbox{ApJ}}
\newcommand{\aap}{\mbox{A\&A}}

\begin{document}
\title*{The European Large Area ISO Survey\protect\newline 
(ELAIS): Latest Results}
\toctitle{The European Large Area ISO Survey\protect\newline 
(ELAIS): Latest Results}
%
%
\titlerunning{ELAIS: Latest Results}
%
\author{Seb  Oliver\inst{1,5}
\and S. Serjeant\inst{1}
\and A. Efstathiou\inst{1}
\and H. Crockett\inst{1}
\and C. Gruppioni\inst{2}
\and F. La Franca\inst{3}
\and M. Rowan-Robinson\inst{1}
\and ELAIS Consortium\inst{4}}
\authorrunning{Seb Oliver et al.}
%
%
\institute{
Astrophysics Group, Blackett Laboratory, Imperial College of 
Science Technology \& Medicine, Prince Consort
Rd., London. SW7 2BW, England\\
\and Osservatorio Astronomico di Bologna, 
 Bologna, Italy\\
\and Dipartimento di Fisica, Universita degli Studi ``Roma TRE'', Roma, Italy\\
\and 23 additional institutes as detailed on http://athena.ph.ic.ac.uk/
\and Astronomy Centre, CPES,  University of Sussex, Brighton, BN1 9QJ, England\\
emailto:S.Oliver@sussex.ac.uk
}
%

\maketitle              

\begin{abstract}
We present some recent results from the European Large Area ISO 
Survey (ELAIS).  This survey was the largest non-serendipitous ISO
field survey.  
A preliminary reduction has recently been completed and catalogues of
sources released to the community.  Early results show strongly
evolving source counts.  A comprehensive identification programme is
underway and a number of extremely luminous objects have already been
discovered.  This survey provides an exciting legacy from the ISO
mission and (amongst many goals) will allow us to provide important
constraints on the obscured star-formation history of the Universe.

\end{abstract}

\section{Introduction}

The European Large Area ISO Survey (ELAIS~\cite{Oliver et al. 2000}) 
was the largest
non-serendipitous ISO~\cite{Kessler et al. 1996} field survey, utilising 377 hours of 
the Open-Time  programme.  The project is a collaborative
venture between 26 
institutes from 11 countries, almost exclusively European.
ELAIS is the major project supported by the TMR network 
programme ``ISO Surveys''.  

The survey covers around 12 square 
degrees and has observations in four bands covering
much of the ISO wavelength window, $6.7, 15, 90$ and $175\umu$m,
using both {\em ISO-CAM\/}~\cite{Cesarsky et al. 1996}
 {\em ISO-PHOT\/}~\cite{Lemke et al. 1996}.  

As with any field survey the goals are varied and 
include the following:
\begin{itemize}
\item{Obscured star formation history of the Universe}
\item{Ultra-luminous Infrared Galaxies at high$-z$}
\item{Dusty tori around AGN}
\item{Dust in Normal Galaxies}
\item{Dust emission from halo stars}
\item{Detection of new classes of objects}
\item{Investigation of populations making up the FIR background}
\end{itemize}
A more detailed description
of the principal intended goals can of the ELAIS survey can be found in 
Oliver et al. 2000~(ELAIS~\cite{Oliver et al. 2000}) 
though pre-defined goals will not anticipate the full range of
possibilities that the data affords.   

\begin{figure}
\centering
\includegraphics[width=.6\textwidth, angle=90]{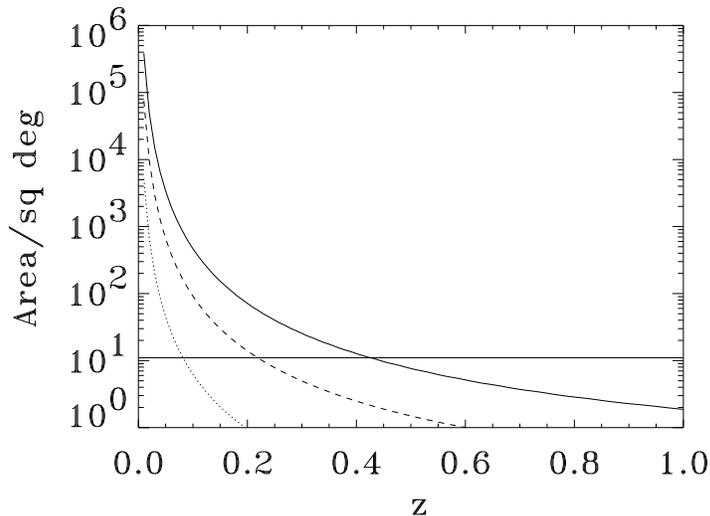}

\caption[]{The minimum area of a survey required to measure mean densities in
populations visible to a given redshift such that the systematic
errors due to large-scale structure are: $\sigma=0.1$ - solid line,
$\sigma=0.2$ - dashed line, $\sigma=0.5$ - dotted line.   The nominal
area of the {\em ELAIS} survey is over-plotted.  This plot assumes
that the survey area is split into four independent survey areas
as is the case for {\em ELAIS}. 
}\label{fig:cosmic_var} 


\end{figure}

For a number of the goals we require statistically significant
samples of galaxies at high redshifts ($z\sim0.5$)  compared to e.g. IRAS but
relatively modest compared to some deep surveys.  
It was this requirement that drove the survey area to be of order
10 square degrees.  Not only does a large area 
produce larger samples of galaxies,  reducing the Poisson errors,
it also reduces the cosmic variance, which would otherwise be
significant at these redshifts.  Fig. \ref{fig:cosmic_var} illustrates
the minimum survey area required to measure a global 
quantity (e.g. mean galaxy density) with a given cosmic variance.

This illustrates that ELAIS  is complimentary to smaller area
deeper surveys ISO such as the HDF \cite{Serjeant et al. 1997}
which study similar populations at higher redshift but cannot
address these populations at lower redshift.

\section{ISO Observation Summary}

ELAIS comprised four major fields N1-3, S1 and 7 smaller fields
S2, X1-6 (three of which were chosen on the basis of known 
objects and so are not true field surveys).  A summary of the area
of each field covered in each band is present in Table \ref{tab:field_areas}

\begin{table}
\centering
\caption[]{Survey Fields covered at least once.   Areas are given in
square degrees.  The 175$\umu$m observations in N1 have been 
carried out by the FIRBACK team (PI J-L Puget, see Dole et
al. \cite{Dole et al. 1999})
and are included in this table to illustrate the complete {\em ISO} coverage
of the {\em ELAIS} fields.  N1-3, S1-2 and X1-3 are unbiased survey
fields, while X4-X6 are centred on known objects so should not be
included with the other fields for statistical purposes.
}
\begin{tabular}{ccccc}
Field & \multicolumn{4}{c}{Wavelength/$\umu$m} \\
     & 6.7     & 15     & 90    & 175 \\
     &       &        &       &     \\
N1   &       &  2.67  &  2.56 & 2    \\
N2   & 2.67  &  2.67  &  2.67 & 1    \\
N3   & 1.32  &  0.88  &  1.76 &     \\
S1   & 1.76  &  3.96  &  3.96 &     \\
S2   & 0.12  &  0.12  &  0.11 &  0.11   \\
X1   &       &  0.16  &  0.19 &     \\
X2   &       &  0.16  &  0.19 &     \\
X3   &       &  0.16  &  0.19 &     \\\hline
     & 5.87  & 10.78  & 11.63 &  3.11 \\ \hline
X4   &       &  0.09  &  0.11  &     \\
X5   &       &  0.09  &        &     \\
X6   &       &  0.09  &  0.11  &     \\
\end{tabular}
\label{tab:field_areas}
\end{table}

\section{ELAIS Data Products}

In order to provide source lists and maps for rapid follow-up we pursued a
two phase strategy for the data processing.  During the mission we 
decided on a ``Preliminary'' data reduction pipe-line.  This processing has
been carried out on the complete survey data, producing source lists both for
follow-up campaigns and for preliminary scientific analysis.  An important
stage in the ``Preliminary'' data reduction for both PHOT and CAM
data was that at least two observers examined the time-lines of candidate
sources before acceptance.  This extremely labour intensive activity ensured
that the sources lists are highly reliable.  Subsets of the ``Preliminary Analysis Catalogues'' were released to world via our WWW page (http://athena.ph.ic.ac.uk/) concurrent with the release of data in August 1999.  
Table \ref{tab:pa} shows the number of sources in subsets of the
catalogues. 

\begin{table}
\centering
\caption[]{Numbers of objects in ``Preliminary'' Analysis catalogues
as released via the WWW and in the entire catalogues. The released
CAM catalogues had a hard flux limit imposed,
approximate limiting fluxes are given for other samples.}\label{tab:pa}

\begin{tabular}{lrrr}
Cat.   & Band  & No. & Flux \\
       & $\umu$m&  &mJy \\
PA-WWW & 7 & 273& 4\\
       &15 &484&	4\\
	&90& 98&	150\\
PA	&15&600&	2\\
	&90&300&60\\

\end{tabular}

\end{table}

The second (``Final Analysis'') phase in
the data reduction, which involves reprocessing the entire data sets, using the best post-mission  knowledge is now nearing completion.  
``Final'' catalogues will be available withing the {\em ELAIS} consortium early in 2000 and to the world  shortly thereafter.

\section{Results from the ISO Data}

The first results to be extracted from the {\em ELAIS} data naturally come from the {\em ISO} data alone.  Much more detailed insight will be forthcoming when the data are combined with follow-up observations and surveys at other 
wavelengths.

\begin{figure*}

\vspace{-2.5cm}
\includegraphics[width=.6\textwidth]{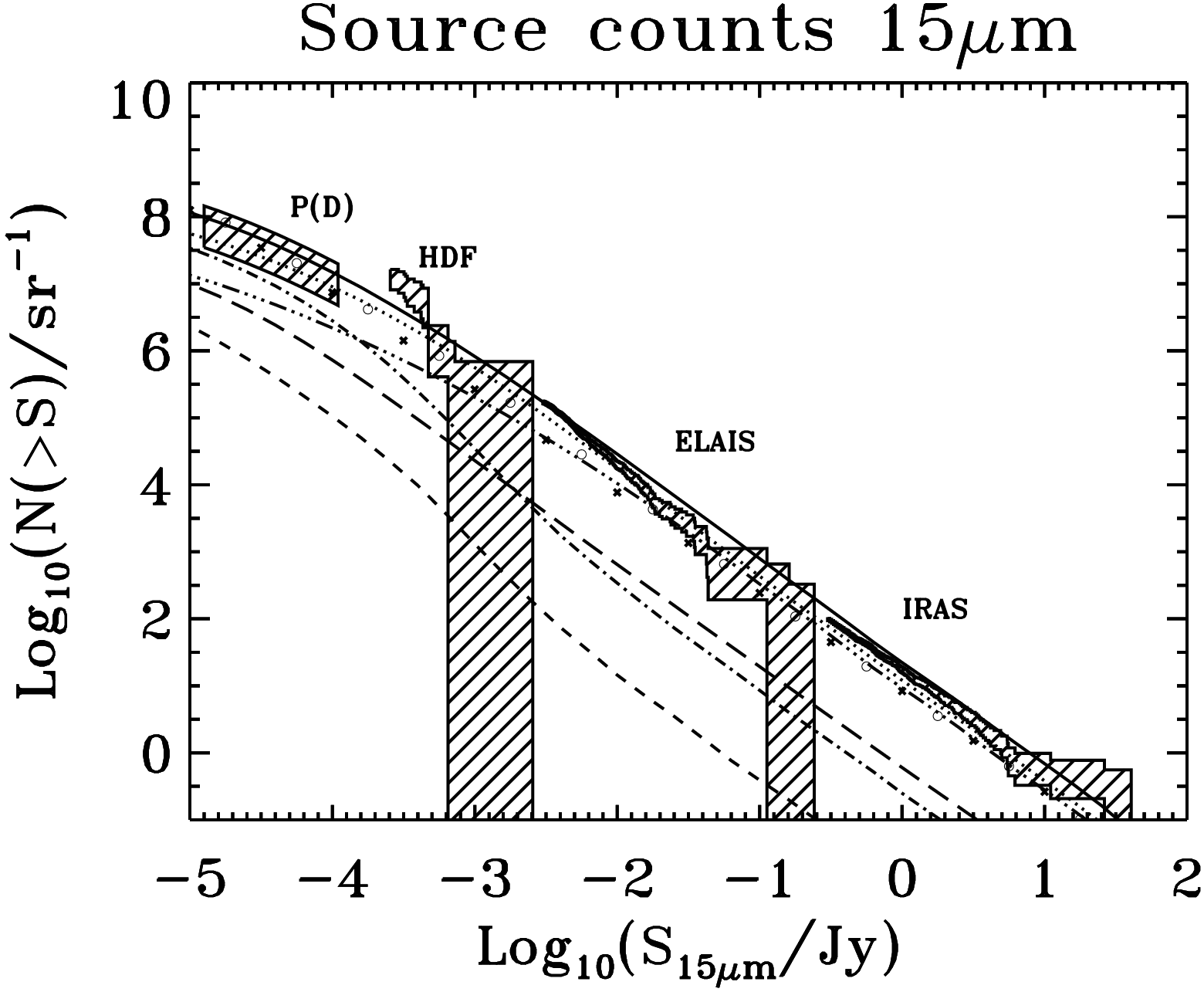}

\vspace{-2.5cm}
\includegraphics[width=.6\textwidth]{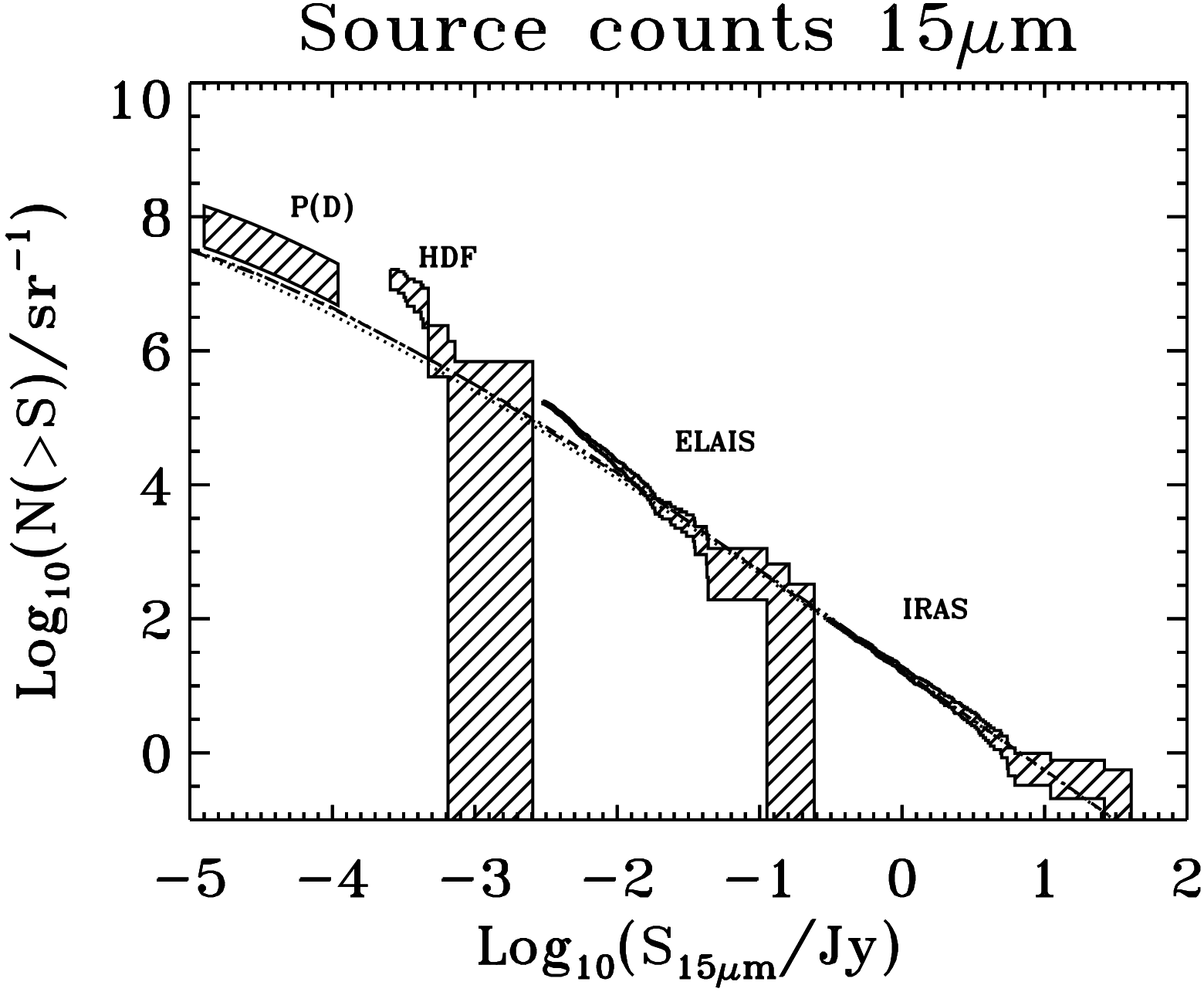}

\vspace{+0.5cm}

\caption[]{
Integral ELAIS extragalactic source counts at $15\umu$m. Flux densities
are quoted in Jy.  
Shaded regions show the ranges spanned by $\pm1\sigma$ uncertainties.
Also shown are the source counts and P(D) analysis from the Hubble
Deep Field \cite{Oliver et al. 1997}, \cite{Aussel et al. 1999}.
The IRAS counts are estimated from
the $12\umu$m counts.

The upper panel has the Franceschini et al. source count model
\cite{Franceschini et al. 1994} over-plotted.  
The model spiral contribution is shown as a dotted line, ellipticals
as a dashed line, S0 as a dash-dot line, star-bursts as a
dash-dot-dot-dot line and AGN as a long dashed line. 
The total population model is shown as a full line. 
Also plotted are the Guiderdoni et al. \cite{Guiderdoni et al. 1998} models A and E, as
small filled crosses and small open circles respectively. 
The lower panel shows all available no-evolution models. 
Franceschini et al. (in prep.) is 
over-plotted as a dash-dot-dot-dot line. 
The Xu et al. \cite{Xu et al. 1998} models are shown with 
and without the MIR spectral features (dash-dot and dotted 
respectively). All three no-evolution models have been renormalised
to match the IRAS counts, by a factor of $0.8$ for the Franceschini et 
al. models, and $1.8$ in the case of the Xu et al. \cite{Xu et al. 1998} models. 
}\label{fig:15counts}
\end{figure*}

\begin{figure}
\centering
\includegraphics[width=.6\textwidth]{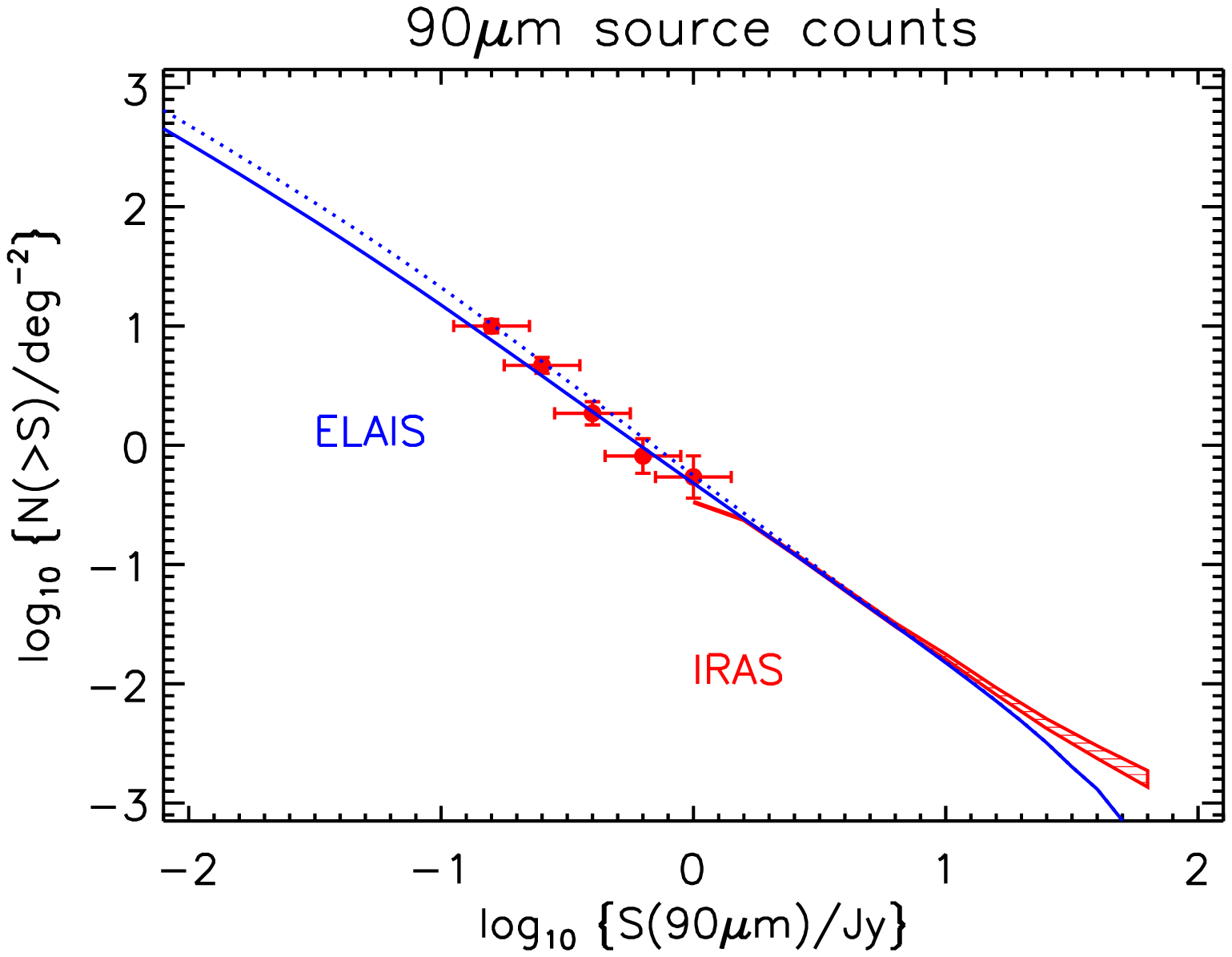}
\caption[]{
{\em ELAIS} and {\em IRAS} 90$\umu$m source counts.  The solid line
and dotted lines are the counts predicted by the models A and E (respectively)
of Guiderdoni et al. \cite{Guiderdoni et al. 1998}
}\label{fig:90counts} 


\end{figure}

\begin{figure}
\includegraphics[width=.6\textwidth,angle=-90]{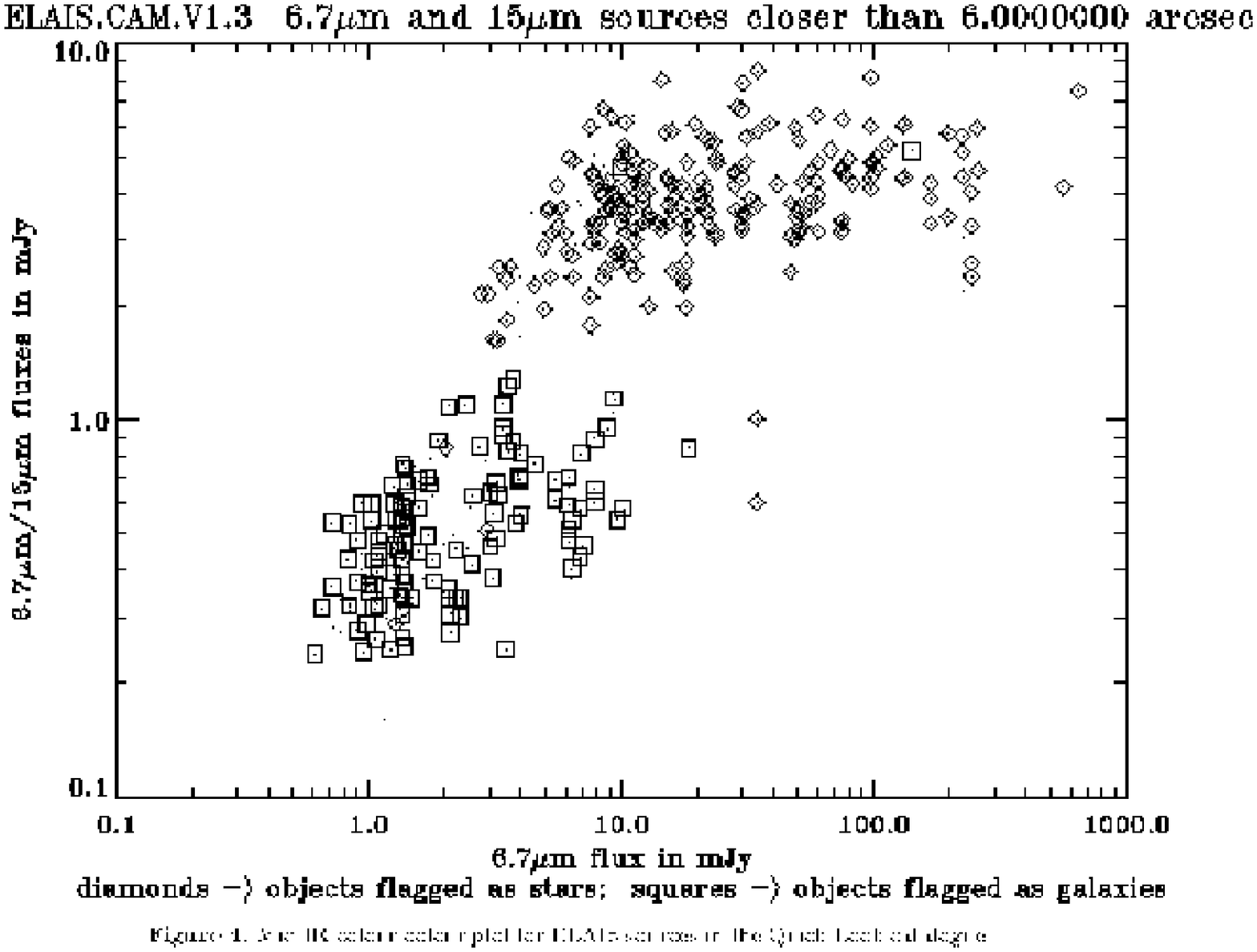}

\caption{Mid-IR colour-colour diagram for ELAIS sources in the Preliminary
Catalogue.  Diamonds indicate sources classified as ``stars'' on the
basis of the morphology of their apparent counterparts on the DSS plates;
squares indicate sources classified as ``galaxies''; isolated dots
indicate sources which were too faint for a clear morphological classification.
}
\label{fig:colours}
\end{figure}

The galaxy number counts have been determined at 6.7, 15 and 90$\umu$m.  
The 15$\umu$m counts\cite{Serjeant et al. 2000}, illustrated in Figure 
\ref{fig:15counts},  have been shown to agree with many strongly evolving 
population models\cite{Franceschini et al. 1994},
\cite{Pearson and Rowan-Robinson 1996}, \cite{Xu et al. 1998}
 based on {\em IRAS} counts,  while being inconsistent with at least one
non-evolving model \cite{Xu et al. 1998}. 
The 90$\umu$m counts \cite{Efstathiou et al. 2000}, Figure \ref{fig:90counts}, are also consistent with similar strong evolution models.
The 6.7 and 15$\umu$m fluxes of morphologically classified 
stars were in general consistent with model photospheres and, 
as expected,
 the  6.7 and 15$\umu$m fluxes provide a good discriminant 
between stars and galaxies, Figure \ref{fig:colours} 
\cite{Crockett et al. 2000}.


\section{Follow-up Programme}

An extensive follow-up programme is being undertaken, utilising a vast 
battery of telescopes around the world, and also a number of satellites.
The current coverage of the {\em ELAIS} fields in un-biased surveys
across the electromagnetic spectrum is summarised in 
Table \ref{tab:fup}.  

The spectroscopic follow-up has not progressed as rapidly as we would have
liked.  The largest sample of spectra we have come from the S1 field taken in 
one hour of 2dF time, snatched from an otherwise cloudy night.  Another
cloudy night on the 2dF produced a second exposure, though this has yet to 
be analysed.

\begin{table}
\begin{centering}
\begin{small}
\caption{Multi-wavelength field surveys within the main {\em ELAIS} fields, the vast majority carried out as part 
of the {\em ELAIS} collaboration.  Areas are in square degrees. 
Some sub-fields within these go to greater depth.  The X-ray and
sub-mm surveys are yet to be completed.
}\label{tab:fup}

\begin{tabular}{lcccccccccc}
Band       & 2-10keV &  $u,g,r,i,z$ &    $H$ & $K$ & 6.7 & 15  & 90 & 175  & 850  & 21cm  \\
Depth Units      & CGI       &   mag    & mag  & mag   & mJy & mJy & mJy & mJy & mJy  & mJy   \\
\\
\multicolumn{11}{c}{N1}\\
Area       & 0.07       &    9       &   0.5 & 0.4 &    & 2.6  & 2.6 &  2   & 0.05 &  1.54  \\
Depth      & $10^{-14}$
                      &   23.3,24.2,23.5,    & 19.5 & 18.0 & 1   & 3   & 100 & 100 & 8 &  0.1-0.4 \\
& & 22.7,21.1 &&&&&&&&\\
\\
\multicolumn{11}{c}{N2}\\
Area       & 0.07       &    2       &  0.5   & 0.4&  2.7  & 2.7  & 2.7 &  1   & 0.05 &  1.54  \\
Depth      & $10^{-14}$     
                  &   22.5,24.2,23.5, & 19.5 & 18.0 & 1   & 3   & 100 & 100 & 8 &  0.1-0.4 \\
& & 22.7,21.1 &&&&&&&&\\
\\
\multicolumn{11}{c}{N3}\\

Area       &        &    1,1,2.3,1,0      &   1 &  &  1.32  & 0.9  & 1.76 &   &  &  1.14 \\
Depth      &     &   22.5,23,23,      & 19.5 & & 1   & 3   & 100 &  &  &  0.1-0.4 \\
\\
& & 23,0 &&&&&&&&\\
\multicolumn{11}{c}{S1}\\

Area       & 2        & 1.2,0,4,3,0      &    &  &  1.8  & 4  & 4 &
&  &  4  \\
Depth      & $10^{-13}$
                      &   23,0,23.5,  & &  & 1   & 3   & 100 &  &  &  0.24 \\
& & 23,0 &&&&&&&&\\

\end{tabular}
\end{small}
\end{centering}

\end{table}

Since the Ringberg meeting two surveys using XMM to study the {\em ELAIS}
fields and a La Palma International Time Programme
have been awarded time.  The XMM surveys will provide both deeper and 
wider hard X-ray coverage than is indicated in Table \ref{tab:fup} (which refers
only to the Chandra and BeppoSax surveys).   The International Time 
is around three weeks on La Palma telescopes, primarily this
will be used to obtain spectra for our {\em ISO} sources,  but will also 
provide deeper optical and NIR imaging.

\section{Early Results from Follow-up Programme}

The limited spectroscopy we have been able to obtain to date has allowed us
to determine redshifts for around 300 {\em ISO} and radio sources. 
A couple of the {\em ISO} sources are at redshifts $z>3$, and many sources
have ultra high luminosities.
From these spectra we are also able to present a very preliminary 
classification, Table \ref{tab:class}.  All the types of objects that we
expected from studies of local {\em IRAS} samples are seen, it is too early
to  say whether the populations are found in the proportions expected as the
selection effects in these highly incomplete spectroscopic samples need to be 
modeled carefully.

\begin{table}
\centering
\caption[]{Provisional spectral classifications of {\em ISO} and 21cm
selected sources in the Southern {\em ELAIS} field S1.  Spectra come
from a one hour 2dF exposure and a number of nights on the ESO 3.6m and
NTT telescopes.}\label{tab:class}
\begin{tabular}{lrrrr}
Class & \multicolumn{2}{c}{21cm}& \multicolumn{2}{c}{\em ISO} \\
      &   2dF        &    ESO      &   2dF    & \multicolumn{1}{c}{ESO} \\
      &              &             &          &    \\
Absorption 	& 48    &   	&15	& 6	\\
Star-burst 	& 9	&3	&22	&52 \\
H$\alpha$	&5	&2	&26	&\\
OII		&10	&	&4	&\\
OIII		&1	&	&	&\\
AGN/QSO		&6	&2	&20	&19\\
AGN/Sy1		&2	&1	&3	&\\
AGN/Sy2		&8	&3	&8	&8\\
AGN/BLLac	&	&	&2	&\\
Stars		&1	&	&8	&3\\
Too Faint	&60	&3	&41	&\\
\end{tabular}
\end{table}

\section{Conclusion}
The {\em ELAIS} sample is clearly going to provide a major legacy from the {\em ISO} mission.
A preliminary catalogue has already been released to the
community and a final analysis is in progress. Source counts show strong
evolution, and with forthcoming optical spectroscopy in 2000 ELAIS will
provide exceptional constraints on the cosmic star formation history.

\clearpage
\addcontentsline{toc}{section}{Index}
\flushbottom
\printindex

\end{document}